\documentclass[fleqn,natbib]{mnras}

\usepackage[T1]{fontenc}
\usepackage{ae,aecompl}

\usepackage{graphicx}	% Including figure files
\usepackage{amsmath}	% Advanced maths commands
\usepackage{amssymb}	% Extra maths symbols

\title[Isochrone models]{Spherical isochrone models revisited}

\author[H. Dejonghe]{
Herwig Dejonghe\thanks{E-mail: Herwig.Dejonghe@UGent.be}
\\
Astronomical Observatory, Ghent University, Krijgslaan 281, B9000 Gent, Belgium
}

\date{Accepted XXX. Received YYY; in original form ZZZ}

\pubyear{2020}

\begin{document}

\label{firstpage}
\pagerange{\pageref{firstpage}--\pageref{lastpage}}
\maketitle

\begin{abstract}
A short derivation of all isochrone models using complex analysis.
\end{abstract}


%%%%%%%%%%%%%%%%%%%%%%%%%%%%%%%%%%%%%%%%%%%%%%%%%%

%%%%%%%%%%%%%%%%% BODY OF PAPER %%%%%%%%%%%%%%%%%%

\section{Introduction}
The study of orbits in spherical potentials is an essential part of any course in stellar dynamics. Usually 3 types of potentials are considered for which the radial period, defined as the time it takes the orbit to go from pericenter to pericenter (or apocenter to apocenter) does not depend on the angular momentum: the quadratic, the Kepler and the isochrone potentials. The first 2 types are quite elementary. The isochrone models on the other hand were discovered by M. H\'enon in 1959. His proof that they exist is brilliant but rather long and tedious. The primary purpose of this paper is to recover these 3 classes of potentials by simpler means, at least if one has complex analysis in the toolbox, while all the way keeping the original characterization of isochrone potentials as the target. Another way to arrive at the isochrone potentials is the requirement that the radial action be expressible in terms of elementary functions (Evans {\it et.al.} 1990, and references therein). 

\section{The analysis}
\label{sect_analysis}
Be $r$ the radial coordinate, $\dot r$ the radial velocity, $V(r)$ the binding potential, $E$ the binding energy and $L$ the modulus of the angular momentum, all per unit mass. These are connected by
\begin{equation}
\frac12 \left({\dot r}^2+{L^2\over r^2}\right) - V(r) = -E.
\end{equation}
The radial period then equals
\begin{equation}
\label{defT}
T = 2\int_{r_p}^{r_a}{r\,dr\over\sqrt{2r^2\left[V(r)-E\right]-L^2}}
\end{equation}
with $r_p<r_a$ the pericenter and apocenter radii. It is essential that they are simple roots of the radicand, by definition of the orbits we are looking for. A positive radicand implies $E\le V(r)$. 

We can see the integral (\ref{defT}) as an integral in the complex plane $z$ of the form
\begin{equation}
\int {\phi(z)\,dz\over\sqrt{(z-a)(b-z)}}=
{1\over i}\int {\phi(z)\,dz\over\sqrt{(z-a)(z-b)}},
\end{equation}
with the real roots $a\le b$ following from the roots $r_p$ and $r_a$ via the transformation $z(r)$. Next we set up 2 branch cuts $]-\infty,a]$ and $]-\infty,b]$, with branch points of the order $(z-a)^{-1/2}$ and $(b-z)^{-1/2}$, yielding effectively a branch cut $[a,b]$. We now consider the contour:
\begin{enumerate}
\item a line from $a$ to $b$ with ${\rm Im}(z)\!\to\!+0$
\item a clockwise circle with radius $R\!\to\!0$ around $b$
\item a line from $b$ to $a$ with ${\rm Im}(z)\!\to\!-0$
\item a clockwise half circle ${\rm Im}(z)\!<\!0$ with radius $R\!\to\!0$ around $a$
\item a line from $a$ to $-\infty$ with ${\rm Im}(z)\!\to\!-0$
\item a counterclockwise circle with radius $R\!\to\!+\infty$, which we denote by $\oint$
\item a line from $-\infty$ to $a$ with ${\rm Im}(z)\!\to\!+0$
\item a clockwise half circle ${\rm Im}(z)\!>\!0$ with radius $R\!\to\!0$ around $a$.
\end{enumerate}
The contributions of the circle around $b$, (ii), and the circle around $a$, (iv) plus (viii), tend to zero. As to the phase of $[(z-a)(z-b)]^{-1/2}$ on the contour lines, we see that (v) and (vii) cancel, since the integrand is holomorphic there, safe for the presence of poles $d_i$, while (i) and (iii) add to $-T$. The residue theorem yields
\begin{eqnarray}
&&\oint {\phi(z)\,dz\over\sqrt{(z-a)(b-z)}}-T =\nonumber\\ 
& &\qquad\qquad+2\pi i\sum_{c_i}{\rm Res}[\phi(z)]
- 2\pi i\sum_{d_i}{\rm Res}[\phi(z)].
\end{eqnarray}
The $c_i$  are the poles, if any, in the complex plane other than the $d_i$ on the real axis smaller than $a$. Hence 
\begin{eqnarray}
T &=& \oint {\phi(z)\,dz\over\sqrt{(z-a)(b-z)}}+\nonumber\\ 
& &\qquad+2\pi i\sum_{d_i}{\rm Res}[\phi(z)]
- 2\pi i\sum_{c_i}{\rm Res}[\phi(z)].
\end{eqnarray}
We must exclude the presence of any other branch cut arising from $\phi(z)$, since it would generate another integral depending explicitly on $L^2$. Hence we are looking for coordinate transformations and/or functions $V(r)$ with the additional requirements that the integrand
\begin{enumerate}
\item should produce a finite result for the contour integral
\item must have only poles as singularities except for the only branch cut, which must be a square root of a second degree polynomial in the integration variable that is positive on the interval concerned.
\end{enumerate}
We begin our quest with the form
\begin{equation}
\label{defoTx}
\int_{r_p^2/2}^{r_a^2/2}{dz\over\sqrt{4z\left[V(z)-E\right]-L^2}}, \qquad z={r^2\over2},
\end{equation}
for part (i) of the contour. We do not consider any additional constant in $V(z)$, which can be trivially absorbed in $E$. Similarly, any potential of the form
\begin{equation}
\label{defW}
W(z) = V(z) - {\lambda\over4z}=V(r)-{\lambda\over2r^2}=W(r)
\end{equation}
will also do, since the additional term can be absorbed in the term $L^2$, replacing effectively $L^2$ by $L^2+\lambda$. 

The only finite potential that satisfies both requirements for the integral as written in (\ref{defoTx}) is the harmonic oscillator
\begin{equation}
\label{defquad}
V=E-a^2z=E-\frac12a^2r^2.
\end{equation}
There are no poles in the complex $z$-plane, but the contour integral itself equals 
\begin{equation}
\label{Tquad}
\oint{dz\over\sqrt{4zV(z)}}=\oint{dz\over2aiz}=
\lim_{R\to+\infty}\int_0^{2\pi}{d(Re^{i\theta})\over2aiRe^{i\theta}}=\frac\pi a.
\end{equation}

The only other case occurs when we interchange the dependent and independent variable in (\ref{defoTx}):
\begin{equation}
\label{defoTV}
\int_{V_p}^{V_a}{dz\over dV}{dV\over\sqrt{4z(V)(V-E)-L^2}}.
\end{equation}
The most general form for $z(V)$ satisfying the requirements is a fraction with a linear function in the numerator and a simple square in the denominator. In this section we will adopt
\begin{equation}
\label{exprxV}
z(V) = {\mu\over2}{\mu-2bV\over V^2} \quad {\rm or}\quad 2zV^2+2b\mu V-\mu^2=0 ,
\end{equation}
featuring two arbitrary constants $\mu$ and $b$. Without losing generality we can set $b\ge0$ while $\mu$ can have both signs. Note that the denominator is sufficiently general since $V$ is determined apart from an additive constant. The particular form of the expression (\ref{exprxV}) is chosen such as to obtain for one of the roots:
\begin{equation}
\label{Henon}
V(z) = {\mu\over b+\sqrt{b^2+2z}}={\mu\over b+\sqrt{b^2+r^2}}=V(r).
\end{equation}
This is, assuming $\mu>0$, H\'enon's isochrone potential. We obtain for (\ref{defoTV}):
\begin{equation}
\label{intiso}
\int_{V_a}^{V_p}{\mu(\mu-bV)\,dV\over V^2\sqrt{(-4\mu b\!-\!L^2)V^2+2\mu(\mu\!+\!2bE)V-2\mu^2E}}.
\end{equation}
Now the contour integral in the complex $V$-plane tends to 0, but the integrand has one second order pole at $V=0$, which is of the type $d_i$. Hence, after calculation,
\begin{equation}
\label{Tgen}
T=2\pi{|\mu|\over(2E)^{3/2}}.
\end{equation}
Clearly $E>0$ in order to have a radial period. When $b\to0$ we recover the Kepler potential $V(r)=\mu/r$. When both $\mu$ and $b$ are sufficiently large we recover the quadratic potential
\begin{equation}
V(r) = {\mu\over2b}\left(1-{r^2\over4b^2}\right).
\end{equation}
Comparison with (\ref{defquad}) yields $E=\mu/(2b)$ and $a^2=\mu/(4b^3)$ or $a=(2E)^{3/2}/(2\mu)$, making the result (\ref{Tquad}) in accordance with (\ref{Tgen}), as should.

We conclude that the expression (\ref{exprxV}) produces the 3 classes `all-in-one', as set out in the introduction.

We end this section with considering the second solution of the quadratic (\ref{exprxV}), which is
\begin{equation}
\label{Henon2}
V(z) = -\mu{b+\sqrt{b^2+r^2}\over r^2}.
\end{equation}
For $\mu>0$ we have $V(z)<0$. The radicand in (\ref{defoTV}) must be positive in order to have motion, and since (\ref{Tgen}) is independent of the choice of the solution of (\ref{exprxV}), $E>0$.  The angular momentum can become arbitrarily positive, therefore no term $L^2+\lambda$ will prohibit initial conditions that do not allow motion. 

When $\mu<0$, on the other hand, the solution (\ref{Henon}) can be rejected for the same reason as above, while (\ref{Henon2}) is a valid one. It can be written as 
\begin{equation}
\label{Henon2a}
V(r) = |\mu|{b+\sqrt{b^2+r^2}\over r^2}={|\mu|\over b+\sqrt{b^2+r^2}} +{2b|\mu|\over r^2},
\end{equation}
and therefore is not really a new solution in view of (\ref{defW}).

\section{More isochrones}

Clearly, there are more potentials that satisfy the isochrone hypothesis, since the ansatz  (\ref{exprxV}) is still not the most general ansatz one can make. Ramon \& Perez (2020, and references therein) offer a classification of all cases. They take into account the $\lambda$-term of (\ref{defW}), and distinguish the harmonic family (\ref{defquad}), the H\'enon and the Kepler family (\ref{Henon}) and the so called bounded family
\begin{equation}
\label{bounded}
V(r) =- {\mu\over b+\sqrt{b^2-r^2}},\qquad \mu<0.
\end{equation}
Indeed, the term in $V^{-2}$ in ansatz (\ref{exprxV}) carries the factor $\mu^2/2$, which is positive, but need not to. Hence we can cover the negative case by putting
\begin{equation}
\label{exprxVa}
z(V) = -{\mu\over2}{\mu+2bV\over V^2}
\quad {\rm or}\quad 2zV^2+2b\mu V+\mu^2=0.
\end{equation}
With this ansatz the integral (\ref{defoTV}) transforms to
\begin{equation}
\label{intisob}
\int_{U_p}^{U_a}{\mu(\mu-bU)\,dU\over U^2\sqrt{(-4\mu b\!-\!L^2)U^2+2\mu(2bE\!-\!\mu)U+2\mu^2E}}
\end{equation}
with $U=-V$. We find 
\begin{equation}
\label{Tgenb}
T=2\pi{|\mu|\over(-2E)^{3/2}}.
\end{equation}
In this case $E<0$ in order to have a radial period. When solving (\ref{exprxVa}) we recover (\ref{bounded}). The other solution is
\begin{equation}
\label{bounded2}
V(r) = \mu{b+\sqrt{b^2-r^2}\over r^2}={\mu\over b+\sqrt{b^2-r^2}} +{2b\mu\over r^2},\quad\mu>0.
\end{equation}
The same considerations as to the sign of $\mu$ can be made as in the H\'enon case. 

We need not further get into the nature of the orbits for all these isochrone potentials, but refer to \cite{RP} for an extensive discussion. 
%We suffice with a brief discussion of the radicand. Two real roots are only possible if
%\begin{equation}
%2|E|(L^2+\lambda)\le(\mu-2b|E|)^2.
%\end{equation}
%The radicand is positive between these roots (and hence enabling motion) when
%\begin{equation}
%L^2>-4\mu b-\lambda.
%\end{equation}

\section{A bonus calculation and a theorem}
We can use the same method of complex integration to calculate the change in polar angle during one radial period, the apsidal angle:
\begin{equation}
\label{deltath}
\Theta = 2L\int_{r_p}^{r_a}{dr\over r\sqrt{2r^2\left[V(r)-E\right]-L^2}},
\end{equation}
or in its complex form, for part (i) of the contour:
\begin{equation}
\label{deltathx}
\Theta = \frac L2\int_{r_p^2/2}^{r_a^2/2}{dz\over z\sqrt{4z\left[V(z)-E\right]-L^2}}.
\end{equation}

With ansatz (\ref{exprxV}) we obtain
\begin{equation}
\label{deltathiso}
\Theta = L\int_{V_a}^{V_p}{{\displaystyle(\mu-bV)\over\displaystyle(\mu-2bV)}\,dV\over \sqrt{(-4\mu b\!-\!L^2\!-\!\lambda)V^2+2\mu(\mu\!+\!2bE)V-2\mu^2E}}.
\end{equation}
Note that this time we included explicitly the term $L^2+\lambda$ instead of $L^2$.  The contribution of the contour integral is $L\pi/\sqrt{4\mu b+L^2+\lambda}$ while the simple pole $V=\mu/(2b)$ of type $c_i$ yields $L\pi/\sqrt{\lambda+L^2}$, making for
\begin{equation}
\label{deltheta}
\Theta=\pi L\left({1\over\sqrt{\lambda+L^2}}+{1
\over\sqrt{4\mu b+\lambda+L^2}}\right),
\end{equation}
valid for the harmonic family ($\mu$ and $b$ large), the H\'enon family and the Kepler family ($b=0$). 

For the bounded case we find similarly, with ansatz (\ref{exprxVa}),
\begin{equation}
\label{delthetab}
\Theta=\pi L\left({1\over\sqrt{\lambda+L^2}}-{1\over\sqrt{4\mu b+\lambda+L^2}}\right).
\end{equation} 
If we would redo the analysis in section \ref{sect_analysis} with the apsidal angle as a goal, it would show that the isochrone potentials are the only ones for which the apsidal angle does not depend on $E$ for bound orbits. 

We thus obtain the characterization of the isochrone orbits as the orbits for which
\begin{enumerate}
\item the radial period does not depend on the angular momentum
\item the apsidal angle does not depend on the energy.
\end{enumerate}

\section*{Acknowledgements}
I like to thank J. Perez to draw my attention to his work and that of his collaborators.

\section*{Data availability}
No data were generated or analyzed in support of this research.

%\appendix

%\section{}

% Don't change these lines
\bsp	% typesetting comment
\label{lastpage}

\begin{thebibliography}{}
%\bibitem[B]{B} Binney, J., eprint arXiv:1411.4937
\bibitem[Evans {\it et.a.} (1990)]{Ev}Evans, N.W., de Zeeuw, P.T. \& Lynden-Bell, D., 1990, MNRAS, 244, 111
\bibitem[H\'enon (1959)]{Ha} H\'enon, M., 1959a, Ann. Ap., 22, 126
\bibitem[Henon]{Hb} H\'enon, M., 1959b, Ann. Ap., 22, 491
\bibitem[RM (2020)]{RP} Ramon, P. \& Perez, J., Cel. Mech. \& Dyn. Astr., 2020, 132, 22 
\end{thebibliography}
\end{document}